\documentclass[aps,prd,amsmath,superscriptaddress,twocolumn,showpacs,captions=nooneline]{revtex4-1} 
\usepackage{graphicx,amsmath,amssymb,bm,color,inputenc,fontenc,mathrsfs}
\usepackage{epstopdf}
\usepackage{float}
\usepackage{slashed}
\usepackage{comment}
\usepackage[caption = false]{subfig}
\usepackage{bm} 
\usepackage{todonotes}
\begin{document}

\title[title]{Rate of dark photon emission from electron positron annihilation in massive stars}
\author{Ermal Rrapaj}
 \email{ermalrrapaj@gmail.com}
\affiliation{School of Physics and Astronomy, University of Minnesota, Minneapolis, MN 55455, USA}
\affiliation{Department of Physics, University of California, Berkeley, CA 94720, USA}
\author{Andre Sieverding}
%\email{asieverd@umn.edu}
\affiliation{School of Physics and Astronomy, University of Minnesota, Minneapolis, MN 55455, USA}
\author{Yong-Zhong Qian}
% \email{qian@physics.umn.edu}
\affiliation{School of Physics and Astronomy, University of Minnesota, Minneapolis, MN 55455, USA}

\begin{abstract}
We calculate the rate of production of dark photons from electron-positron pair annihilation in hot and dense matter characteristic of
supernova progenitors. Given the non-linear dependence of the emission rate on the dark photon mass 
and current astrophysical constraints on the dark photon parameter space, we focus on the mass range of 1--10 MeV.
For the conditions under consideration both mixing with the in-medium photon and plasma effects on the electron dispersion relation are non-negligible and are explored in detail.
We perform our calculations to the leading order in the fine-structure constant.
Transverse and longitudinal photon modes are treated separately given their different dispersion relations.  
We consider the implications for the evolution of massive stars when dark photons decay 
either into particles of the standard model or of the dark sector. 

\end{abstract}
\maketitle
\section{Introduction}
\label{section:intro}
Observations of galaxy rotation curves, the motion of galaxies in clusters, the $\Lambda$CDM model of the universe, and empirical evidence from the bullet cluster 
are among the many indicators of the existence of dark matter (DM) that interacts with ordinary matter through gravitational interactions~\cite{Olive:2003iq,Freese:2008, Roots2010}. 
Many proposed DM models also naturally predict non-gravitational interactions. 
Here we focus on a class of these models, in which DM is part of a neutral hidden sector. This sector interacts with standard model (SM) particles through the exchange of light vector bosons that couple to SM conserved currents \cite{Holdom:1986,Rajpoot:1989,Nelson:1989fx,Batell:2014yra}.  
Dark matter is charged under a local $U(1)$ symmetry in which the mediator couples to the SM electric charge $Q$, and is described by the spin-one field $A^{*}_\mu$, called the dark photon, which mixes kinetically with the standard photon $A_\mu$ \cite{Jaeckel:2010ni}.

The Lagrangian of this system is
\begin{equation*}
 \begin{split}
 \mathcal{L}= -\frac{1}{4}A_{\mu\nu}A^{\mu\nu}-\frac{1}{4}A^{*}_{\mu\nu}{A^{*}}^{\mu\nu}-\frac{\epsilon}{2}A^{*}_{\mu\nu}A^{\mu\nu}+\frac{1}{2}m^2_{A^{*}} A^{*}_\mu {A^{*}}^{\mu} ,
 \label{eq:lagrangian}
 \end{split}
\end{equation*}
which also includes the mass term for the gauge boson and where $A_{\mu\nu}=\partial_{\mu}A_{\nu}-\partial_{\nu}A_{\mu},\ A^{*}_{\mu\nu}=\partial_{\mu}A^{*}_{\nu}-\partial_{\nu}A^{*}_{\mu}$. 
The emission and absorption of dark photons are due to the oscillations between these two vector bosons. In the stellar medium, both transverse and longitudinal photon modes exist and both mix with the dark photon. Because they have different dispersion relations, they behave differently and we study both.
Plasma effects in the kinetic mixing have been considered previously  both in the existing constraints on the dark photons from SN 1987A and from our sun \cite{Redondo:2013,Rrapaj:2015,Chang:2016,Hardy:2017}. 
The purpose of this work is to re-visit the emission rates and mean free path for the dark photon for density and temperature ranges relevant for the evolution of massive stars ($ \gtrsim 8 M_{\odot}$) and study possible implications
for the unconstrained parameter space.

In the core of a supernova, nucleon-nucleon bremsstrahlung is the dominant process of $A^{*}$ production. However, for pre-supernova massive stars, pair annihilation is highly favored as stars get sufficiently hot while still at densities far below those of a proto-neutron
star. Under these conditions we expect pair annihilation to be the major emission channel of $A^{*}$. Bremsstrahlung among nuclei is a sub-leading source of dark photon emission and will be considered in future investigations.

Here, we systematically include medium effects in calculating dark photon emission from electron positron pair annihilation in massive stars and compare the rates with 
those of neutrino emission.
In particular, we provide a complete consistent treatment of plasma effects on photon and electron dispersion relations. Given that the dark photon is thermally produced, we focus on the mass range 1--10 MeV, 
which is relevant for the temperatures inside massive stars, and is unconstrained by supernova considerations \cite{Chang:2016,Hardy:2017}.
We aim to provide a self-contained treatment that can serve as a reference for numerical implementation in stellar evolution. In addition, we qualitatively discuss how the dark photon emission might affect massive star evolution.

We start by analysing the mixing of the dark photon with the standard photon in section \ref{section:mix}.
In section \ref{section:pair} we focus on dark photon emission from electron-positron pair annihilation. 
In section \ref{section:compare} we compare dark photon emission with neutrino emission for conditions typical of massive stars. We also consider how the dark photon can affect the evolution
of these stars. We intend to employ the rates provided here in stellar evolution calculations in subsequent works to quantify the effects of the dark photon.

\section{Dark photon production}

\subsection{Mixing between dark photons and photons}
\label{section:mix}
To study the mixing of the two gauge bosons, the dark photon field is redefined as $B^{\mu}= {A^{*}}^{\mu} - \epsilon A^{\mu}$, which transforms the kinetic mixing into a mass mixing:
\begin{equation*}
 \begin{split}
  \mathcal{L}= -\frac{1}{4}A_{\mu\nu}A^{\mu\nu}-\frac{1}{4}B_{\mu\nu}B^{\mu\nu}+\frac{1}{2}m^2_{A^{*}}( B_\mu -\epsilon A^{\mu})^2.
 \end{split}
\end{equation*}
Following the same procedure as in \cite{Redondo:2013}, we perform a perturbative expansion in $\epsilon$. While production of $B_{\mu}$ is of order $\epsilon^{2}$, the difference between $B^{\mu}$ and the original dark photon
 ${A^{*}}^{\mu}$ is of order $\epsilon^4$, which for practical purposes is negligible (we assume $\epsilon \leq 10^{-9}$). 
 The decay width $\Gamma$ and the imaginary part of the polarization tensor $\Pi$ \cite{Weldon1983} are related by
 \begin{equation}
  \text{Im}[\Pi]=-\omega \Gamma=-\omega(\Gamma_{\text{abs}}-\Gamma_{\text{ems}}),
 \end{equation}
 where $\omega$ is the dark photon energy. By employing detailed balance, $\Gamma_{\text{abs}}=e^{\omega/T}\Gamma_{\text{ems}}$, the  emission rate can be expressed as
 \begin{equation*}
 \begin{split}
  \Gamma_{\text{ems}}=&-\frac{\text{Im}[\Pi]}{\omega (e^{\omega/T}-1)}.
 \end{split}
 \end{equation*}
 \begin{figure}[htbp]
 \includegraphics[width=0.75\linewidth]{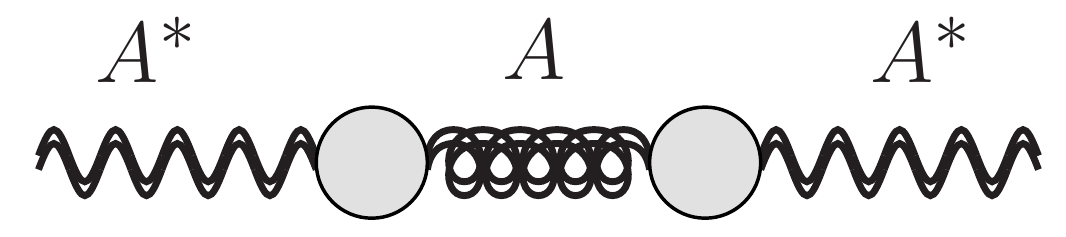}
 \caption{Mixing between dark and standard photon at order $\mathcal{O}(\epsilon^2)$}
 \label{fig:mix}
\end{figure}

Figure~\ref{fig:mix} shows the leading order diagram of the mixing of the two types of photons, which gives
\begin{equation*}
 \begin{split}
  \Pi_{A^{*}}=&m^2_{A^{*}} + \frac{\epsilon^2 m^4_{A^{*}}}{\omega^2-k^2-\Pi_{A}}+\mathcal{O}(\epsilon^4).
 \end{split}
\end{equation*}
The dark photon emission rate is
\begin{equation}
 \begin{split}
  &\Gamma_{A^{*}}{}_{\text{ems}}%=-\frac{\text{Im}[\Pi^{A^{*}}]}{\omega(e^{\omega/T}-1)} \\
  %=&-\frac{\epsilon^2 m^4_{A^{*}}}{\omega(e^{\omega/T}-1)} \frac{\text{Im}[\Pi_{A}]}{(\omega^2-k^2-\text{Re}[\Pi_{A}])^2+\text{Im}[\Pi_{A}]^2}+\mathcal{O}(\epsilon^4)\\
  =\frac{\epsilon^2 m^4_{A^{*}} \ \Gamma_{A}{}_{\text{ems}}}{(\omega^2-k^2-\text{Re}[\Pi_{A}])^2+\omega^2(e^{\omega/T}-1)^2(\Gamma_{A}{}_{\text{ems}})^2}.%+\mathcal{O}(\epsilon^4)
 \end{split}
 \label{eq:gamma}
\end{equation}
The real part of the photon polarization tensor is calculated at one loop level in perturbation theory \cite{Braaten:1993} and is different for transverse and longitudinal modes:
\begin{equation}
 \begin{split}
  \text{Re}[\Pi_{A}^T]=&m_T^2,\\
  \text{Re}[\Pi_{A}^L]=&\frac{\omega^2-k^2}{\omega^2}\omega_L^2=\frac{m_{A^{*}}^2}{\omega^2}\omega_L^2.\\
 \end{split}
 \label{eq:mtwl}
\end{equation}

The thermal mass for the transverse mode $m_T$ and plasma frequency for the longitudinal mode $\omega_L$ are calculated based on the expressions found in~\cite{Braaten:1993}. 
As the dark photon oscillates into a virtual photon, the virtual photon has the same dispersion 
relation as the dark photon, $\omega^2-k^2=m_{A^{*}}^2$.
The transverse thermal mass is
\begin{equation}
 \begin{split}
  m_T^2=&\frac{4\alpha}{\pi}\int_0^{\infty} dk_e \ \frac{k_e^2}{E_e}(f_{e^{-}}+f_{e^{+}}),\\
  \end{split}
  \label{eq:mt}
  \end{equation}
  where $f_{e^{\pm}}$ are the Fermi-Dirac distribution functions for electrons and positrons. 
For most of the evolution of a massive star, the photon thermal mass is less than 1 MeV and is mainly a function of density.
This mass changes drastically only during core collapse,  quickly reaching $\sim$ 14 MeV.
Thus, $m_{A^{*}} \gg m_T$ during most of the stellar evolution. Nevertheless, we employ Eq.~(\ref{eq:mt}) in numerical calculations without any approximation.

While $m_T$ depends only on the properties of the medium, i.e., electron density and temperature, the longitudinal plasma frequency $\omega_L$ is an implicit function of the photon momentum $k$. For momentum values less than
\begin{equation*}
k_{\text{max}}=\omega_P\sqrt{\frac{3}{v_{*}^2}\left[\frac{1}{2v_{*}}\log\left(\frac{1+v_{*}}{1-v_{*}}\right)-1\right]}\,,    
\end{equation*}
$\omega_L$ is the solution
of the following transcendental equation,
  \begin{equation}
  \begin{split}
  \omega_L=&\frac{2 v_{*} k}{3\omega_P^2}\ \frac{v_*^2k^2+3 \omega_P^2}{\log(\frac{\omega_L+v_*k}{\omega_L-v_*k}) -1 }.\\
  \end{split}
  \label{eq:wl}
  \end{equation}
   For $k>k_{\text{max}}$, $\omega_L=k$.
  The plasma frequency $\omega_P$ and effective electron speed $v_*$ are given as:
  \begin{equation*}
  \begin{split} 
  \omega_P^2=& \frac{4 \alpha}{\pi}\int dE_e \frac{k_e^2}{E_e} (1-\frac{1}{3} \frac{k_e^2}{E_e^2})(f_{e^{-}}+f_{e^{+}}),\\
  \omega_1^2=& \frac{4 \alpha}{\pi}\int dE_e \frac{k_e^2}{E_e} (\frac{5}{3} \frac{k_e^2}{E_e^2} -  \frac{k^4}{E_e^4})(f_{e^{-}}+f_{e^{+}}),\\
  v_*=&\frac{\omega_1}{\omega_P}.\\ 
 \end{split}
\end{equation*}
In the high density or high temperature limit, $v_* \sim$ 1. Unlike $m_T$, because $\omega_L$ is momentum dependent, it is not easy to judge the regions of density and temperature
where $\omega_L$ is smaller or larger than the dark photon mass. 
In our calculations, we always solve the transcendental equation in order to perform the integrations required for 
emissivities. For further details on the photon dispersion relation in medium we refer the reader to \cite{Braaten:1993}. 

From Eqs.~(\ref{eq:gamma}) and (\ref{eq:mtwl}) the emission rates for the two modes are
\begin{equation}
 \begin{split}
  \Gamma_{A^{*}}^T{}_{\text{ems}}=&\frac{\epsilon^2 m_{A^{*}}^4 \ \Gamma_A^T{}_{\text{ems}}}{(m_{A^{*}}^2-m_T^2)^2+ \omega^2 (e^{\omega/T}-1)^2 (\Gamma^T_A{}_{\text{ems}})^2},\\
  \Gamma_{A^{*}}^L{}_{\text{ems}}=& \frac{\epsilon^2 m_{A^{*}}^4\ \omega^4\ \Gamma_A^L{}_{\text{ems}}}{m_{A^{*}}^4(\omega^2-\omega_L^2)^2+ \omega^6 (e^{\omega/T}-1)^2 (\Gamma^L_A{}_{\text{ems}})^2}.\\
 \label{eq:mixing}
 \end{split}
\end{equation}
It is interesting to note that for a small dark photon mass, the emission rate for the transverse mode is inversely proportional to that of the photon, whereas for a large dark photon mass, the two rates are proportional.

\subsection{Dark photon emission from $e^{-}e^{+}$ annihilation}
\label{section:pair}
\begin{figure}[htbp]
 \includegraphics[width=0.75\linewidth]{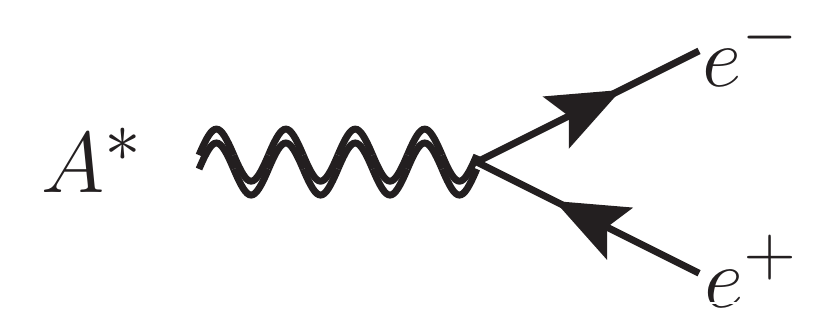}
 \caption{(Dark) virtual photon decay into electron positron pair}
 \label{fig:pairdecay}
\end{figure}

We calculate the absorption rate and obtain the emission rate by multiplying 
the former with the detailed balance factor $e^{-\omega/T}$. 
As the virtual photon decays into standard model particles, here we employ the dark photon dispersion relation. We emphasize that  $A^{*}$ in Fig.~\ref{fig:pairdecay} is the virtual photon on the dark photon mass shell (a real photon cannot decay into $e^{-}e^{+}$ \cite{Braaten:1993}).
The decay rate \cite{Weldon1983} is
\begin{equation*}
 \begin{split}
  \Gamma_{A^{*}\rightarrow e^{-}e^{+}}=& \frac{1}{2 \omega} \frac{1}{4\pi} \int d\Omega_{\omega}  \int \frac{d^3p_{e^{-}}}{2E_{e^{-}}(2\pi)^3} \int \frac{d^3p_{e^{+}}}{2E_{e^{+}}(2\pi)^3}\\
  &| \mathcal{M}_{A^{*} \rightarrow e^{-} e^{+}}|^2 (1-f_{e^{-}})(1-f_{e^{+}})\\
  & \times (2 \pi)^4 \delta^{(4)}(K-P_{e^{-}}-P_{e^{+}}).
 \end{split}
\end{equation*}
As the transverse and longitudinal photon modes behave differently in medium, we calculate the squared matrix element for each mode. The detailed expressions are given in Appendix~\ref{sec:append}.

Plasma effects on the electron and positron dispersion relations are included to be consistent with the correction to the photon dispersion relation described in section~\ref{section:mix}.
In particular, the electron mass gets a correction of order $\alpha^{1/2}$~ \cite{Braaten:1992},
\begin{equation*}
\begin{split}
 \tilde{m}_e=&m_e/2+\sqrt{m_e^2/4+m_{\text{eff}}^2}\,,\\
  m_{\text{eff}}=&\sqrt{\frac{\alpha}{2}(\pi^2T^2+\mu_e^2)}.
\end{split}
\end{equation*}
This effect was included in~\cite{Redondo:2013}, and as that work focused on a smaller dark photon mass relevant for solar conditions, pair annihilation was not considered.
In this work we use $\tilde{m}_e$ in the relevant expressions [see Eq.~(\ref{eq:ratesee})]. This mass can be as high as 12 MeV in a supernova core \cite{Hardy:2017}.

\section{Implications for stellar evolution}

\subsection{Comparison with neutrino emission}
\label{section:compare}
\begin{figure}[ht]
 \includegraphics[width=\linewidth]{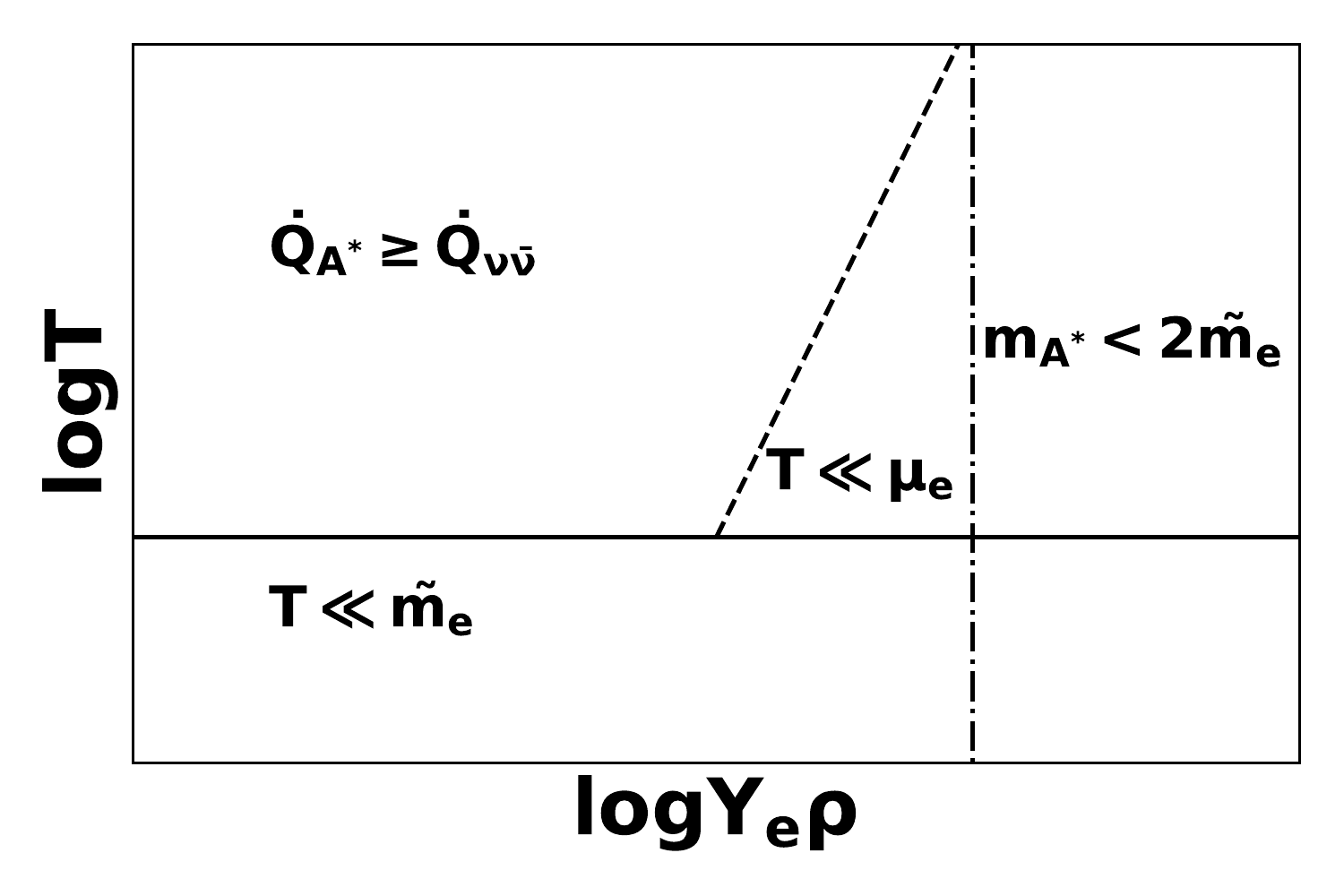}
 \caption{Conditions regarding dark photon emission. Region below solid line: temperature is very low and there are very few electron-positron pairs. 
 Region to the right of the dashed line: density is very high and there are few positrons even when temperature is high.
 Region to the right of the vertical dot-dashed line: dark photon mass is lower than the mass of the pair in the medium.
 Only for the region in the upper left corner is dark photon emission expected to potentially exceed neutrino emission.}
 \label{fig:baseline}
\end{figure}

The rate of total energy emitted from dark photons is
\begin{equation*}
\begin{split}
 \dot{Q}_{A^{*}}=& \int \frac{d^3 k}{(2\pi)^3} \omega \Gamma_{A^{*}}{}_{\text{ems}}.\\
\end{split}
 \end{equation*}
 We use the polarization averaged rate,
 \begin{equation*}
     \Gamma_{A^{*}}{}_{\text{ems}}= \frac{1}{3}\left(\Gamma_{A^{*}}^T{}_{\text{ems}}+\Gamma_{A^{*}}^L{}_{\text{ems}}\right).
 \end{equation*}
Neutrino emission through pair annihilation, photo-neutrino process, plasmon decay, bremsstrahlung, 
and recombination has been studied in detail and standard results are available in the literature \cite{Itoh:1985,Itoh:1986,Itoh:1989,Itoh:1992,Itoh:1996,Haft:1994,Braaten:1993}.
Using these standard neutrino cooling rates we compute the evolution for stars with initial masses of 12, 25, and 30 $M_{\odot}$
and with solar metallicity using the same stellar evolution code as in \cite{WOOSLEY:2007}.
For the conditions covered by these evolutionary tracks, i.e., core temperature and density, we compare the net energy loss rates due to dark photons and neutrinos.
Before performing a quantitative comparison we provide a qualitative assessment of which stellar conditions favor the dark photon emission.
In Fig.~\ref{fig:baseline} we identify three regions where we expect dark photon emission to be suppressed due to low temperatures 
(below the solid line), high density (to the right of the dashed line), and small dark photon mass (to the right of the dot-dashed line), respectively. For relatively hot and not too dense environments we can expect dark photon emission to compete with neutrino emission.
 
\begin{figure}[htbp]
\subfloat[$m_{A^{*}}=1.21$ MeV.]{\includegraphics[height=0.36\linewidth,width=0.48\linewidth]{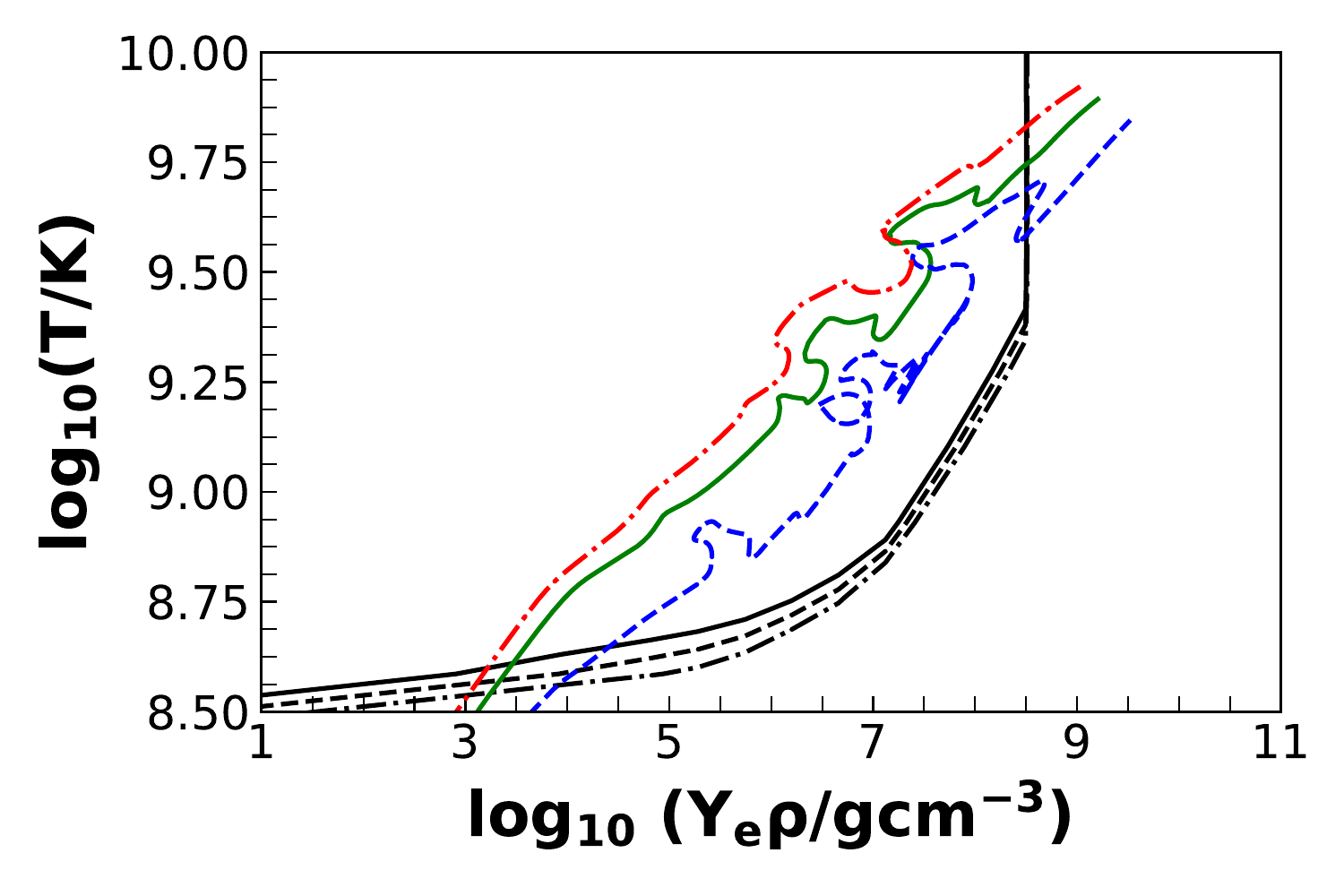}}
 \subfloat[$m_{A^{*}}=2$ MeV.]{\includegraphics[height=0.36\linewidth,width=0.55\linewidth]{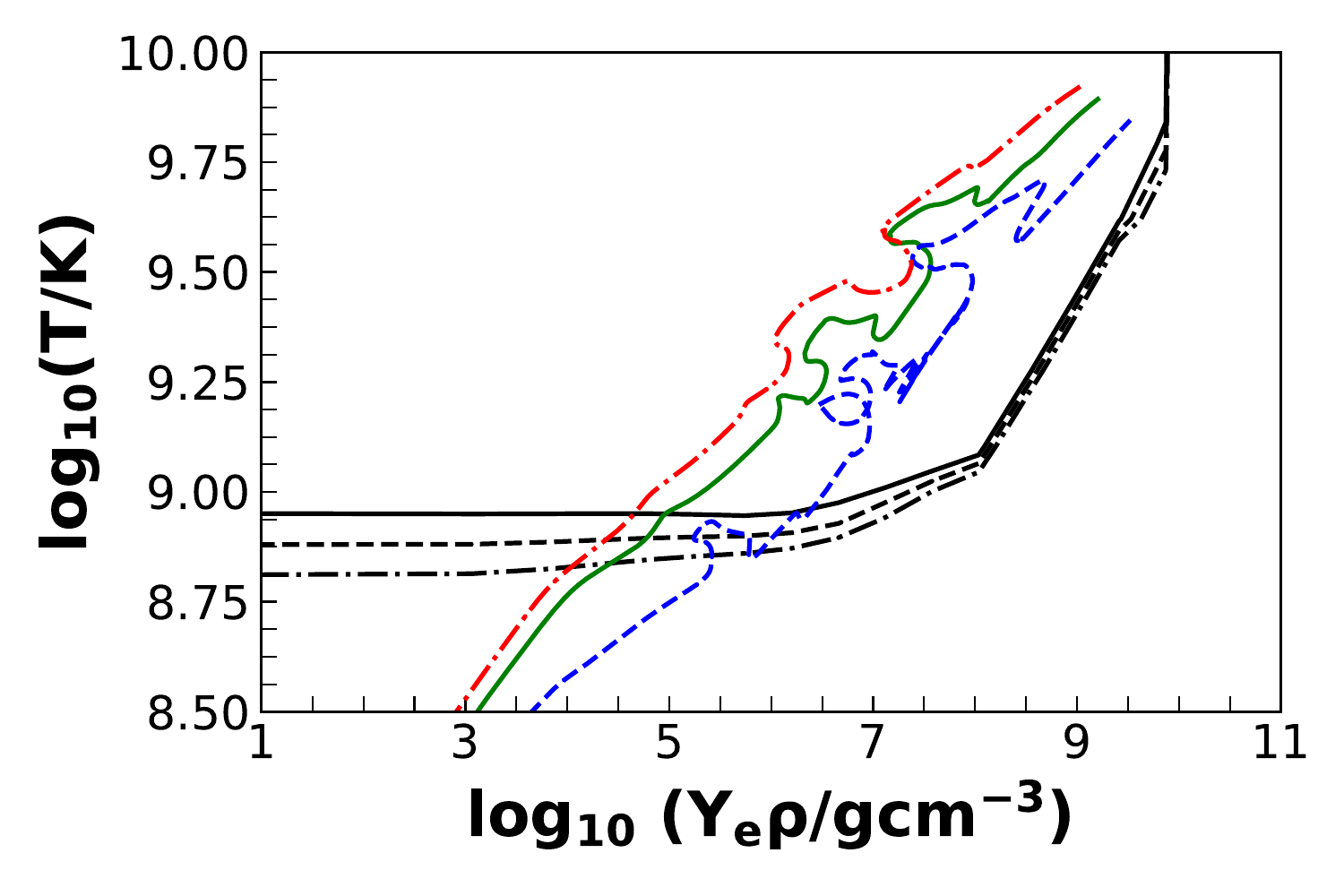}}\\
 \subfloat[$m_{A^{*}}=3$ MeV.]{\includegraphics[height=0.36\linewidth,width=0.48\linewidth]{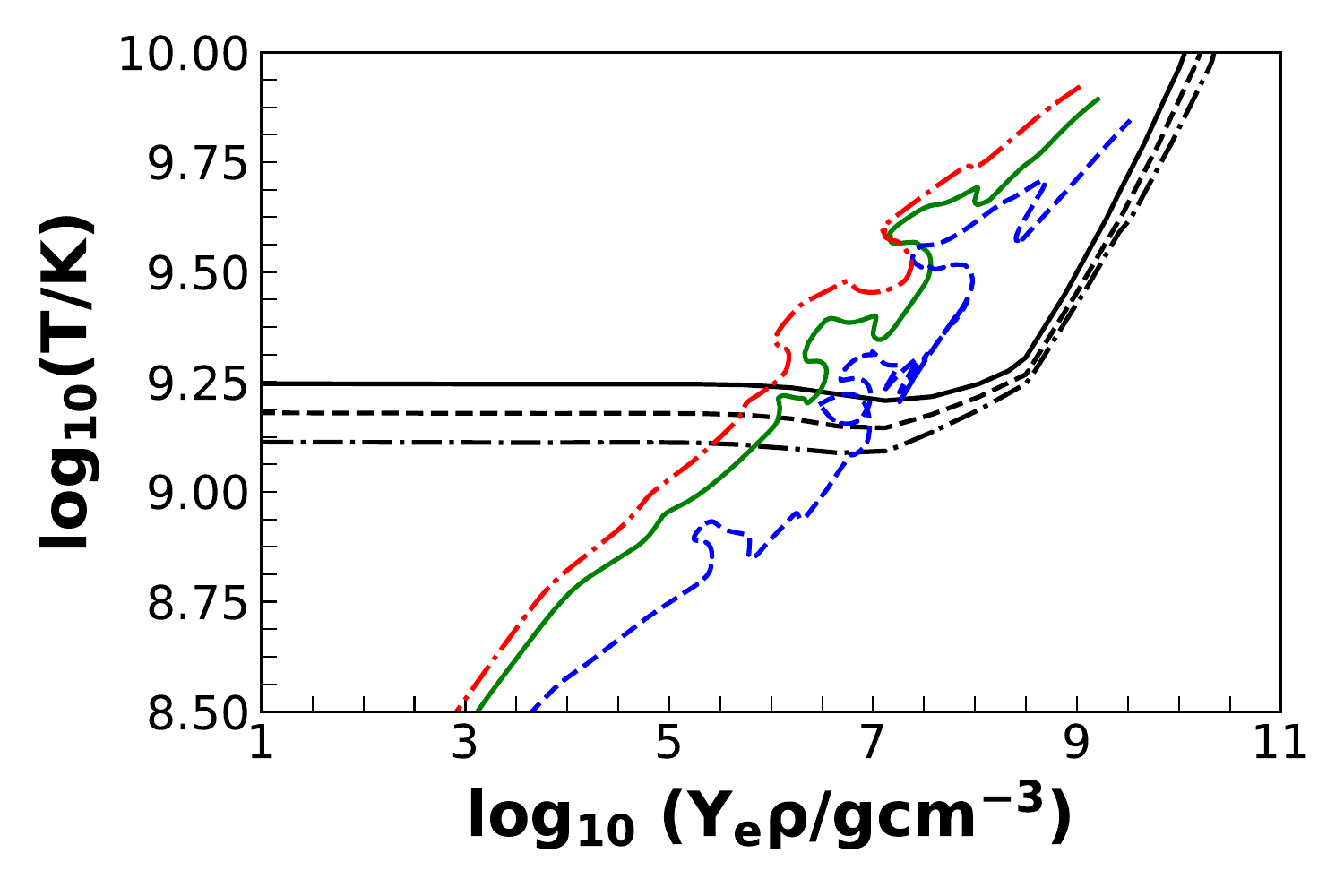}}
 \subfloat[$m_{A^{*}}=4$ MeV.]{\includegraphics[height=0.36\linewidth,width=0.55\linewidth]{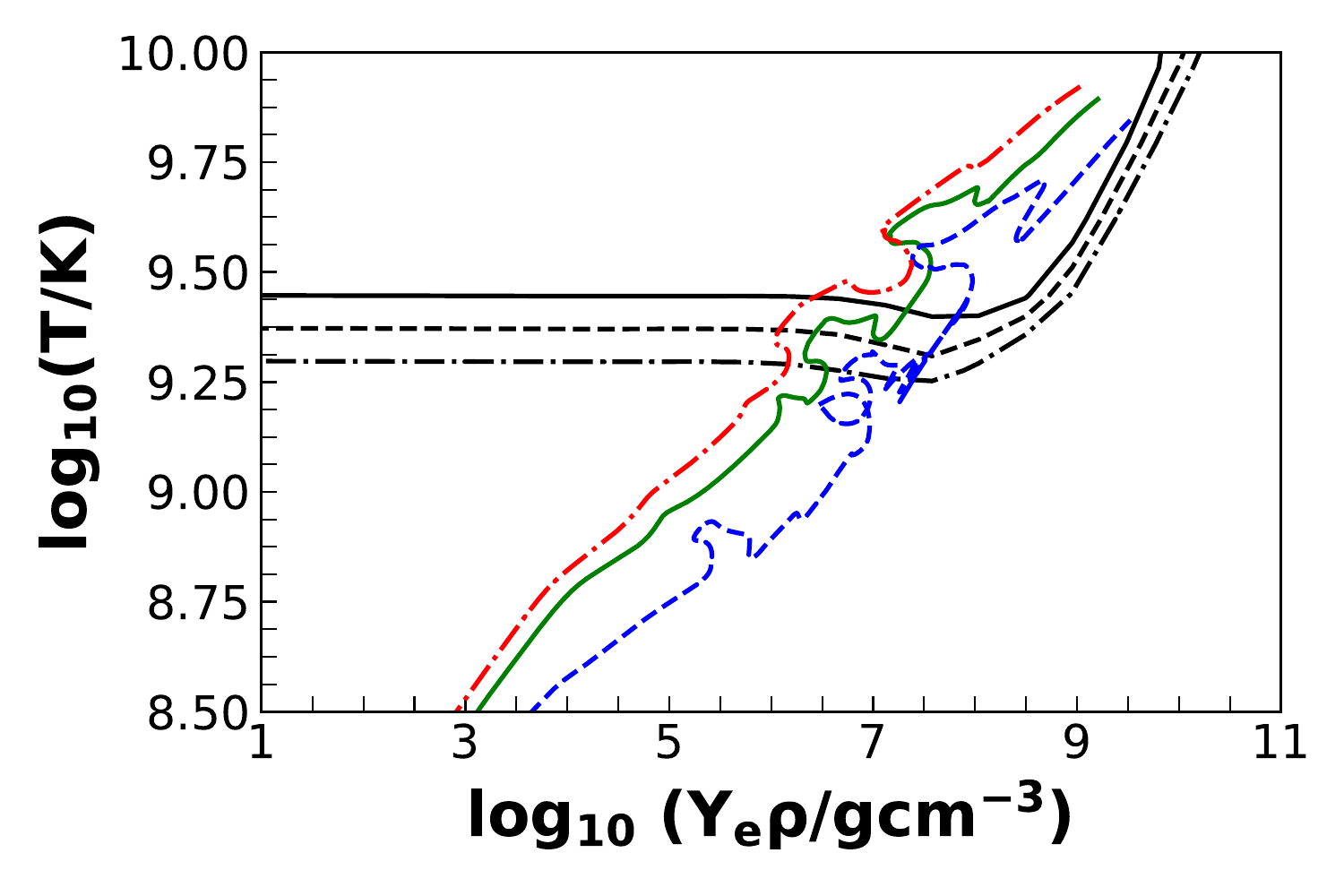}}\\
 \subfloat[$m_{A^{*}}=5$ MeV.]{\includegraphics[height=0.36\linewidth,width=0.48\linewidth]{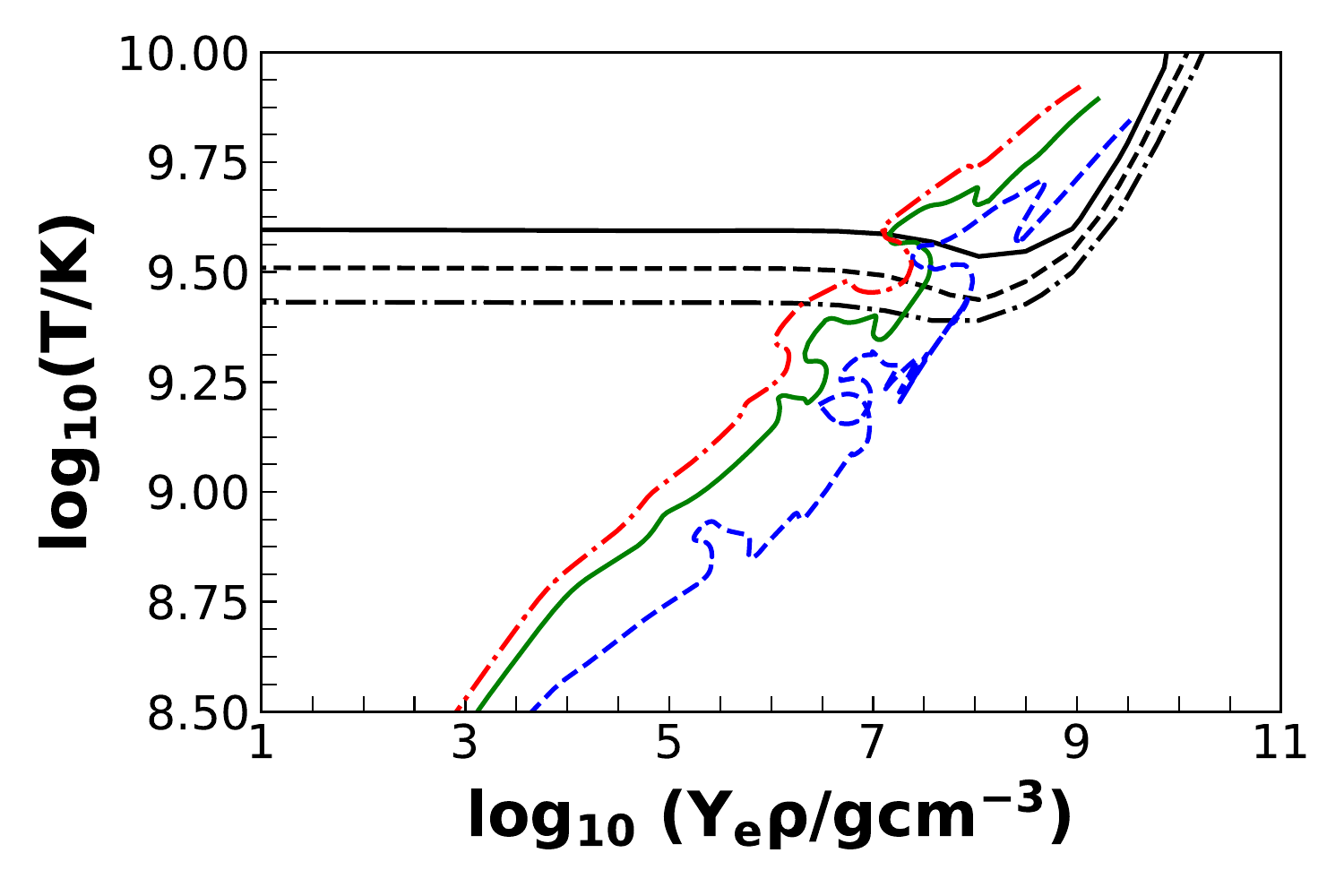}}
 \subfloat[$m_{A^{*}}=10$ MeV.]{\includegraphics[height=0.36\linewidth,width=0.55\linewidth]{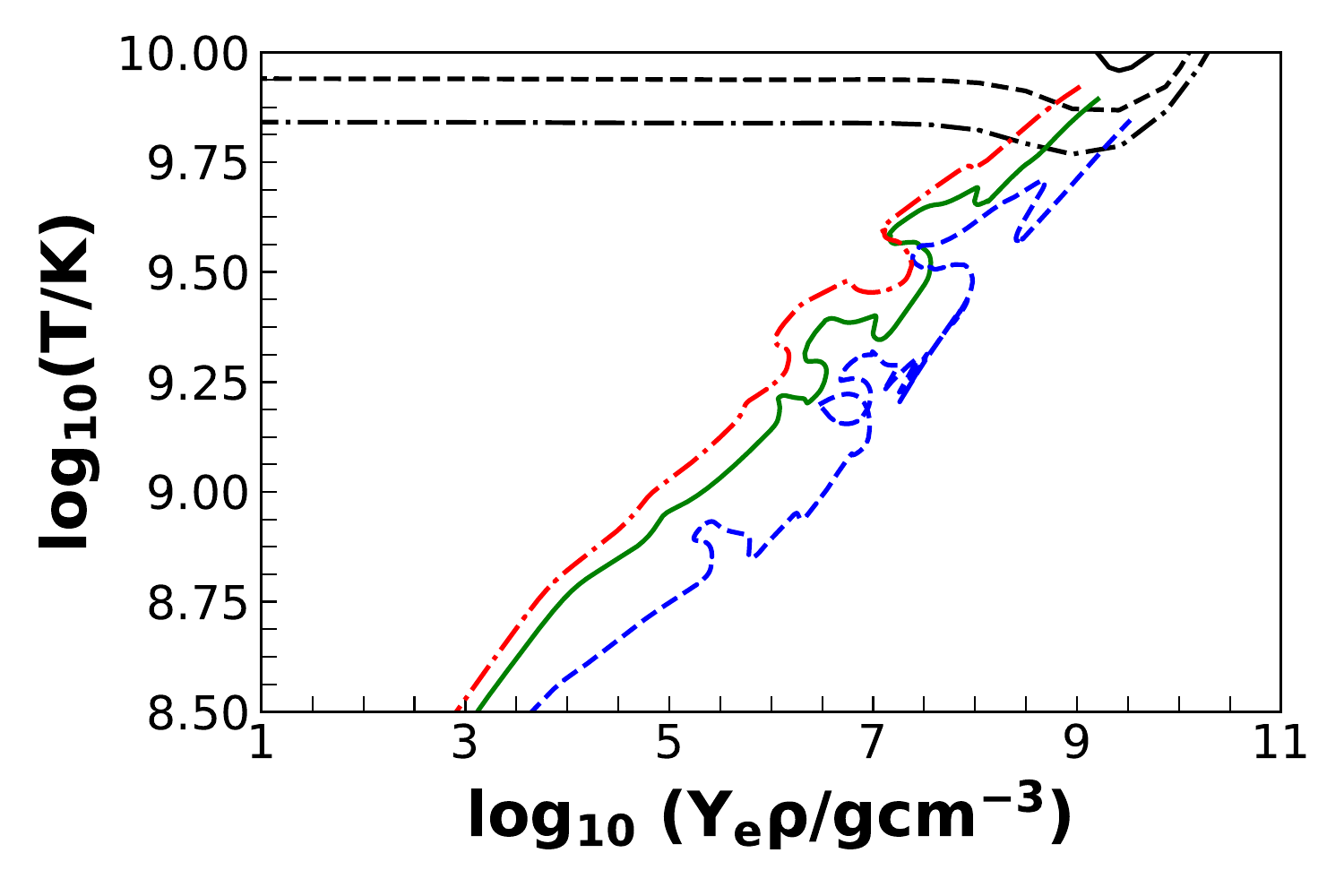}}\\
 \caption{Ratio of dark photon emissivity from pair annihilation to neutrino emissivity as function of electron density and temperature assuming  the reduced coupling strength is $\epsilon=10^{-9}$.
 Stellar evolutionary tracks for a 30 $M_{\odot}$ (red dot-dashed), 25 $M_{\odot}$ (green solid), and 12 $M_{\odot}$ (blue dashed) are also shown. The contour lines represent ratios of 10 (solid), 1 (dashed) and 0.1 (dot-dashed).}
 \label{fig:Eam9_1}
\end{figure}

In Fig.~\ref{fig:Eam9_1}, we show the ratio of the energy loss rates $\dot{Q}_{A^*}/\dot{Q}_{\nu \bar{\nu}}$ for the coupling strength $\varepsilon=10^{-9}$  in the dark photon mass range 1--10 MeV. The stellar evolutionary tracks calculated using neutrino cooling only are also shown in the same figure.
There is a high sensitivity to the dark photon mass for the range 1--5 MeV.
In the 5--10 MeV range, the rates are similar and become relevant at higher temperatures for higher masses. 
In all the cases shown in Fig.~\ref{fig:Eam9_1}, dark photon emission becomes relevant for relatively high temperatures and not very high densities, in agreement with the qualitative assessment shown in Fig.~\ref{fig:baseline}.
Indeed, for smaller dark photon masses we see the expected suppression when electrons in medium become too massive.
Figure~\ref{fig:Eam9_1} shows that the dark photon emission would dominate over the neutrino emission during the advanced stellar burning phases. For the conditions reached in the stellar core the dark photon energy loss rate can exceed the neutrino losses by several orders of magnitude. We therefore expect dark photons in the mass range of  1--10 MeV to affect the last stages of stellar evolution. In particular, for small mass values of 1--3 MeV, the dark photon emission starts to overtake neutrino emission already towards the end of the carbon burning phase.

\subsection{Dark photon decay}
The way in which the dark photon can affect the stellar environment depends crucially on what happens after the particle is emitted. If the dark photon is not a portal to the dark sector, but merely one more particle yet to be discovered, and if its mass is larger than that of the electron-positron pair, it decays back into such a pair and is reabsorbed by standard matter.
In this scenario, from an astrophysical perspective, it could be understood as a short-lived particle that transports 
energy from the inner region of a star to the outer regions. 

However, if the dark photon is an actual portal to the dark sector, its impact on stellar evolution is quite radical. It could decay, 
for instance, into a pair of dark fermions, and this decay may dominate over its re-absorption by standard model particles. 
This scenario has been explored in electron beam dumps for experimental constraints \cite{Izaguirre:2013,Batell:2014}. 
For stellar evolution, dark photons would act as a cooling agent in this scenario.
If there is emission, there must also be absorption. From the rates calculated in the previous sections, the mean free path 
for dark photon absorption is given by $\lambda^{-1}=\Gamma_{A^{*}\rightarrow e^{-}e^{+}}/v_{\omega}$. 
For typical stellar conditions, electrons and positrons are non-degenerate as densities are not too high. As the pre-supernova temperature reaches maximal values 
of $\sim 0.1$ MeV, and dark photons are thermally produced, we expect them to be non-relativistic with a typical momentum value $\sim \sqrt{2 m_A^* T}$. 
Under these conditions, decay into a standard electron-positron pair can be approximated by decay in vacuum and the mean free path is given by
\begin{equation}
\begin{split}
\lambda^{-1}_{A^{*}\rightarrow e^{-} e^{+}}=&\alpha \epsilon^2 \frac{(m_{A^*}^2+2m_{e}^2)}{3k}\sqrt{1-\frac{4m_{e}^2}{m_{A^{*}}^2}}\,.
\end{split}
\label{eq:mfpaee}
\end{equation}
From Eq.~(\ref{eq:mfpaee}) and the stellar conditions we deduce a mean free path of the order of $10^8$--$10^{9}$ cm for a reduced coupling strength $\epsilon=10^{-9}$.
Were this the dominant decay mode, dark photons would indeed act as a mechanism for redistributing energy from the central part of a star to regions of lower density.
This effect has been studied in core collapse supernovae to further constrain the parameter space \cite{Sung2019}.

However, if the dark photon could decay into a pair of dark fermions ($X \bar{X}$), this channel might be much more efficient. As a consequence, the supernova constraint from energy redistribution would not apply in this case.
As an illustration, we can assume $m_X\sim m_e$ and the coupling of the dark photon to dark fermions  (dark electron-positron pair) to be the same as the standard QED coupling.
The ratio of the two mean free paths would be $\lambda_{A^{*}\rightarrow X \bar{X}}/\lambda_{A^{*}\rightarrow e^{-} e^{+}} \sim \epsilon^2 =10^{-18}$. 
Once produced, the dark photon would essentially always 
decay into the dark sector. In this scenario, it is effectively an additional cooling mechanism.
In future works we plan to explore the effects of this additional cooling on stellar evolution in detail.

\section{Conclusions}
\label{sec:Conclusions}
We have calculated the dark photon emission rates for both transverse and longitudinal modes by taking into consideration all plasma effects at leading order in the fine structure constant. 
We have compared the energy loss rates for dark photon emission from electron-positron annihilation with standard neutrino cooling in massive stars
and found that the former can easily overtake the latter for dark photon masses in the range 1--4 MeV.
If the dark photon is a portal to the dark sector, it becomes an additional cooling mechanism.
We intend to investigate the detailed effects of the combined
cooling by dark photons and neutrinos on stellar evolution in follow-up work. 

\begin{acknowledgments}
We thank Sam McDermott for discussion. We also acknowledge Alexander Heger for providing KEPLER, the hydrodynamics code for stellar evolution,
and for helping set up the stellar evolution calculations. 
This work was supported in part by the NSF (PHY-1630782), the Heising-Simons Foundation (2017-228), and the DOE (DE-FG02-87ER40328).  
\end{acknowledgments}

\appendix
\onecolumngrid
\section{Transverse and longitudinal dark photon production from electron--positron pair annihilation}
\label{sec:append}
To obtain the rates for these two modes we need to decompose the scattering matrix using projection operators
\begin{equation*}
 \begin{split}
  &P^T_{i j}=(\delta_{ij}-\frac{q_iq_j}{q^2}),\ P^T_{00}=P^T_{0\mu}=P^T_{\mu 0}=0,\\
  &P^{L}_{\mu \nu}=(\frac{k^2}{m_{A^{*}}^2},\frac{\omega^2}{m_{A^{*}}^2} \frac{k^i k^j}{k^2}),
 \end{split}
\end{equation*}
where $P^{L}$ is three-longitudinal and $P^{T}$ is three-transverse to the momentum $\vec{k}$ of the photon \cite{Weldon:1982}.
The squared scattering matrix element for each mode is
\begin{equation*}
\begin{split}
 | \mathcal{M}^T_{A^{*}\rightarrow e^{-} e^{+}}|^2=&e^2 \epsilon^2\sum_{\lambda} \epsilon^{T}_{\mu}(q,\lambda)\overline{\epsilon}^{T}_{\nu}(q,\lambda) \big[(\slashed{p}_2+\tilde{m}_e)\gamma^{\mu}(\slashed{p}_3
 -\tilde{m}_e)\gamma^{\nu}\big]\\
  =&4 e^2 \epsilon^2 P^{T}_{\mu \nu}\big[ (p_2^{\mu} p_3^{\nu} + p_2^{\nu} p_3^{\mu} - (\tilde{m}_e^2 + p_2\cdot p_3) g^{\mu \nu})\big]\\
  =&  32 \pi \alpha \epsilon^2 \big[ \tilde{m}_e^2 + p_{e^{-}}\cdot p_{e^{+}}- \frac{(\vec{p}_{e^{-}}\cdot \vec{k})(\vec{p}_{e^{+}}\cdot\vec{k})}{k^2} + (\vec{p}_{e^-} \cdot \vec{p}_{e^+}) \big],\\
  | \mathcal{M}^L_{A^{*}\rightarrow e^{-} e^{+}}|^2=&e^2 \epsilon^2 \sum_{\lambda} \epsilon^{L}_{\mu}(q,\lambda)\overline{\epsilon}^{L}_{\nu}(q,\lambda) \big[(\slashed{p}_2+\tilde{m}_e)\gamma^{\mu}(\slashed{p}_3
 -\tilde{m}_e)\gamma^{\nu}\big]\\
  =&4 e^2 \epsilon^2 P^{L}_{\mu \nu}\big[ (p_2^{\mu} p_3^{\nu} + p_2^{\nu} p_3^{\mu} - (\tilde{m}_e^2 + p_2\cdot p_3) g^{\mu \nu})\big]\\
  =&  32 \pi \alpha \epsilon^2 \big( \frac{k^2}{m_{A^{*}}^2}E_{e^{-}}E_{e^{+}}+\frac{\omega^2}{m_{A^{*}}^2}\frac{(\vec{p}_{e^{-}} \cdot \vec{k})(\vec{p}_{e^{+}} \cdot \vec{k})}{k^2}
  -\frac{\omega}{m_{A^{*}}^2}[E_{e^{-}}(\vec{p}_{e^{+}} \cdot \vec{k})+E_{e^{+}}(\vec{p}_{e^{-}} \cdot \vec{k})] +\frac{\tilde{m}_e^2+p_{e^{-}}\cdot p_{e^{+}}}{2}\big).
\end{split}
\end{equation*}
%\twocolumngrid
By employing energy-momentum conservation, the above results can be simplified as
\begin{equation*}
 \begin{split}
  \frac{| \mathcal{M}^T_{A^{*} \rightarrow e^{-} e^{+}}|^2}{32 \pi \alpha \epsilon^2}=& \tilde{m}_e^2+\frac{m_{A^{*}}^2}{k^2} \big[ \frac{1}{4}(\omega^2 + k^2) - \omega E_{e^{-}}+E_{e^{-}}^2\big], \\
  \frac{| \mathcal{M}^L_{A^{*} \rightarrow e^{-} e^{+}}|^2}{32 \pi \alpha \epsilon^2}=& \frac{m_{A^{*}}^2}{k^2}\big( -\frac{1}{4}m_{A^{*}}^2 + \omega E_{e^{-}}- E^2_{e^{-}}\big).
 \end{split}
\end{equation*}

The integration over the phase-space can be performed analytically to give
\begin{equation}
 \begin{split}
 F_n(k,m_{A^{*}},\mu_e,T)=&\mathcal{F}_n(\omega,\mu_e,T,E_{e^{-}}^{+})-\mathcal{F}_n(\omega,\mu_e,T,E_{e^{-}}^{-}),
 \label{eq:phasespace}
 \end{split}
\end{equation}
where
\begin{equation*}
 \begin{split}
 \mathcal{F}_n(\omega,\mu_e,T,x) =& \int d x\ x^n \big[1-f(x,\mu_e,T)\big] \times \ \big[1-f(\omega-x,-\mu_e,T)\big],\\
  \omega=&\sqrt{k^2+m_{A^{*}}^2}\,,\\
  E_{e^{-}}^{\pm}=&\sqrt{\tilde{m}_e^2+\frac{1}{4}\bigg(\omega \sqrt{1-4\frac{\tilde{m}_e^2}{m_{A^{*}}^2}}\pm k\bigg)^2}\,.
 \end{split}
\end{equation*}
Equation~(\ref{eq:phasespace}) can be expressed analytically for all values of $n$. 
%\onecolumngrid
The first two expressions are
\begin{equation*}
 \begin{split}
  \mathcal{F}_0(\omega,\mu_e,T,x)=&\frac{T}{1-e^{-\omega/T}} \bigg[\log\left(\frac{1 + e^{\frac{x - \mu_e}{t}}}{1 + e^{\frac{\omega + \mu-x}{T}}}\right) -x\bigg],\\
  \mathcal{F}_1(\omega,\mu_e,T,x)=&\frac{T}{1-e^{-\omega/T}} \bigg[\log\left(\frac{1+e^{\frac{x - \mu_e}{t}}}{1+e^{-\frac{\omega - x + \mu_e}{T}}}\right) +T\ \text{Li}_{2}(-e^{\frac{x - \mu_e}{T}})-T \ \text{Li}_{2}(-e^{\frac{\omega-x +\mu_e}{T}})\bigg],
 \end{split}
\end{equation*}
where $\text{Li}_n(x)$ is the poly-logarithmic function of order $n$ of $x$.
Expressions for higher values of $n$ can be obtained in a similar fashion.

Finally, the emission rates can be written as
\begin{equation}
 \begin{split}
%\Gamma_{A^{*} \rightarrow e^{-} e^{+}}(\omega,k)=&\frac{\alpha}{3 \omega k } (m_{A^{*}}^2+2m_e^2) F_0(k,m_{A^{*}},\mu_e,T)\\
&\Gamma^T_{ e^{-} e^{+} \rightarrow A^{*}}(\omega,k)= e^{-\omega/T} \frac{2\alpha m_{A^*}^2}{\omega k^3} \epsilon^2 \sum_{n=0}^2 c^T_n F_n(k,m_{A^{*}},\mu_e,T),\\
&\Gamma^L_{e^{-} e^{+} \rightarrow  A^{*}}(\omega,k)= e^{-\omega/T} \frac{2 \alpha m_{A^*}^2}{\omega k^3} \epsilon^2 \sum_{n=0}^2 c^L_n F_n(k,m_{A^{*}},\mu_e,T),\\
\label{eq:ratesee}
\end{split}
\end{equation}
where
\begin{equation}
\begin{split}
&c^T_0=\frac{k^2}{m_{A^*}^2}\tilde{m}_e^2+ \frac{1}{4} (\omega^2 +k^2),\ c^T_1=- \omega , \ c^T_2=1,\\
&c^L_0 = -\frac{1}{4}m_{A^{*}}^2, \ c^L_1=\omega ,\ \ c^L_2=-1.\\
\end{split}
 \end{equation}
We have verified that these expressions agree with those in \cite{Chang:2016}.
\twocolumngrid

%%%%%%%%%%%%%%%%%%%%%


\begin{thebibliography}{26}
\expandafter\ifx\csname natexlab\endcsname\relax\def\natexlab#1{#1}\fi
\expandafter\ifx\csname bibnamefont\endcsname\relax
  \def\bibnamefont#1{#1}\fi
\expandafter\ifx\csname bibfnamefont\endcsname\relax
  \def\bibfnamefont#1{#1}\fi
\expandafter\ifx\csname citenamefont\endcsname\relax
  \def\citenamefont#1{#1}\fi
\expandafter\ifx\csname url\endcsname\relax
  \def\url#1{\texttt{#1}}\fi
\expandafter\ifx\csname urlprefix\endcsname\relax\def\urlprefix{URL }\fi
\providecommand{\bibinfo}[2]{#2}
\providecommand{\eprint}[2][]{\url{#2}}

\bibitem[{\citenamefont{Olive}(2003)}]{Olive:2003iq}
\bibinfo{author}{\bibfnamefont{K.~A.} \bibnamefont{Olive}}, in
  \emph{\bibinfo{booktitle}{{Particle physics and cosmology: The quest for
  physics beyond the standard model(s).}}} (\bibinfo{year}{2003}), pp.
  \bibinfo{pages}{797--851}, \eprint{astro-ph/0301505}.

\bibitem[{\citenamefont{Freese}(2009)}]{Freese:2008}
\bibinfo{author}{\bibfnamefont{K.}~\bibnamefont{Freese}}, \bibinfo{journal}{EAS
  Publ. Ser.} \textbf{\bibinfo{volume}{36}}, \bibinfo{pages}{113}
  (\bibinfo{year}{2009}), \eprint{0812.4005}.

\bibitem[{\citenamefont{{Roos}}(2010)}]{Roots2010}
\bibinfo{author}{\bibfnamefont{M.}~\bibnamefont{{Roos}}},
  \bibinfo{journal}{arXiv e-prints} \bibinfo{eid}{arXiv:1001.0316}
  (\bibinfo{year}{2010}), \eprint{1001.0316}.

\bibitem[{\citenamefont{Holdom}(1986)}]{Holdom:1986}
\bibinfo{author}{\bibfnamefont{B.}~\bibnamefont{Holdom}},
  \bibinfo{journal}{Physics Letters B} \textbf{\bibinfo{volume}{166}},
  \bibinfo{pages}{196 } (\bibinfo{year}{1986}), ISSN \bibinfo{issn}{0370-2693}.

\bibitem[{\citenamefont{Rajpoot}(1989)}]{Rajpoot:1989}
\bibinfo{author}{\bibfnamefont{S.}~\bibnamefont{Rajpoot}},
  \bibinfo{journal}{Phys. Rev. D} \textbf{\bibinfo{volume}{40}},
  \bibinfo{pages}{2421} (\bibinfo{year}{1989}).

\bibitem[{\citenamefont{Nelson and Tetradis}(1989)}]{Nelson:1989fx}
\bibinfo{author}{\bibfnamefont{A.~E.} \bibnamefont{Nelson}} \bibnamefont{and}
  \bibinfo{author}{\bibfnamefont{N.}~\bibnamefont{Tetradis}},
  \bibinfo{journal}{Phys. Lett.} \textbf{\bibinfo{volume}{B221}},
  \bibinfo{pages}{80} (\bibinfo{year}{1989}).

\bibitem[{\citenamefont{Batell et~al.}(2014{\natexlab{a}})\citenamefont{Batell,
  deNiverville, McKeen, Pospelov, and Ritz}}]{Batell:2014yra}
\bibinfo{author}{\bibfnamefont{B.}~\bibnamefont{Batell}},
  \bibinfo{author}{\bibfnamefont{P.}~\bibnamefont{deNiverville}},
  \bibinfo{author}{\bibfnamefont{D.}~\bibnamefont{McKeen}},
  \bibinfo{author}{\bibfnamefont{M.}~\bibnamefont{Pospelov}}, \bibnamefont{and}
  \bibinfo{author}{\bibfnamefont{A.}~\bibnamefont{Ritz}},
  \bibinfo{journal}{Phys. Rev.} \textbf{\bibinfo{volume}{D90}},
  \bibinfo{pages}{115014} (\bibinfo{year}{2014}{\natexlab{a}}).

\bibitem[{\citenamefont{Jaeckel and Ringwald}(2010)}]{Jaeckel:2010ni}
\bibinfo{author}{\bibfnamefont{J.}~\bibnamefont{Jaeckel}} \bibnamefont{and}
  \bibinfo{author}{\bibfnamefont{A.}~\bibnamefont{Ringwald}},
  \bibinfo{journal}{Ann. Rev. Nucl. Part. Sci.} \textbf{\bibinfo{volume}{60}},
  \bibinfo{pages}{405} (\bibinfo{year}{2010}).

\bibitem[{\citenamefont{Redondo and Raffelt}(2013)}]{Redondo:2013}
\bibinfo{author}{\bibfnamefont{J.}~\bibnamefont{Redondo}} \bibnamefont{and}
  \bibinfo{author}{\bibfnamefont{G.}~\bibnamefont{Raffelt}},
  \bibinfo{journal}{JCAP} \textbf{\bibinfo{volume}{1308}}, \bibinfo{pages}{034}
  (\bibinfo{year}{2013}), \eprint{1305.2920}.

\bibitem[{\citenamefont{Rrapaj and Reddy}(2016)}]{Rrapaj:2015}
\bibinfo{author}{\bibfnamefont{E.}~\bibnamefont{Rrapaj}} \bibnamefont{and}
  \bibinfo{author}{\bibfnamefont{S.}~\bibnamefont{Reddy}},
  \bibinfo{journal}{Phys. Rev.} \textbf{\bibinfo{volume}{C94}},
  \bibinfo{pages}{045805} (\bibinfo{year}{2016}), \eprint{1511.09136}.

\bibitem[{\citenamefont{Chang et~al.}(2017)\citenamefont{Chang, Essig, and
  McDermott}}]{Chang:2016}
\bibinfo{author}{\bibfnamefont{J.~H.} \bibnamefont{Chang}},
  \bibinfo{author}{\bibfnamefont{R.}~\bibnamefont{Essig}}, \bibnamefont{and}
  \bibinfo{author}{\bibfnamefont{S.~D.} \bibnamefont{McDermott}},
  \bibinfo{journal}{JHEP} \textbf{\bibinfo{volume}{01}}, \bibinfo{pages}{107}
  (\bibinfo{year}{2017}), \eprint{1611.03864}.

\bibitem[{\citenamefont{Hardy and Lasenby}(2017)}]{Hardy:2017}
\bibinfo{author}{\bibfnamefont{E.}~\bibnamefont{Hardy}} \bibnamefont{and}
  \bibinfo{author}{\bibfnamefont{R.}~\bibnamefont{Lasenby}},
  \bibinfo{journal}{JHEP} \textbf{\bibinfo{volume}{02}}, \bibinfo{pages}{033}
  (\bibinfo{year}{2017}), \eprint{1611.05852}.

\bibitem[{\citenamefont{Weldon}(1983)}]{Weldon1983}
\bibinfo{author}{\bibfnamefont{H.~A.} \bibnamefont{Weldon}},
  \bibinfo{journal}{Phys. Rev. D} \textbf{\bibinfo{volume}{28}},
  \bibinfo{pages}{2007} (\bibinfo{year}{1983}).

\bibitem[{\citenamefont{Braaten and Segel}(1993)}]{Braaten:1993}
\bibinfo{author}{\bibfnamefont{E.}~\bibnamefont{Braaten}} \bibnamefont{and}
  \bibinfo{author}{\bibfnamefont{D.}~\bibnamefont{Segel}},
  \bibinfo{journal}{Phys. Rev.} \textbf{\bibinfo{volume}{D48}},
  \bibinfo{pages}{1478} (\bibinfo{year}{1993}), \eprint{hep-ph/9302213}.

\bibitem[{\citenamefont{{Braaten}}(1992)}]{Braaten:1992}
\bibinfo{author}{\bibfnamefont{E.}~\bibnamefont{{Braaten}}},
  \bibinfo{journal}{\apj} \textbf{\bibinfo{volume}{392}}, \bibinfo{pages}{70}
  (\bibinfo{year}{1992}).

\bibitem[{\citenamefont{{Munakata} et~al.}(1985)\citenamefont{{Munakata},
  {Kohyama}, and {Itoh}}}]{Itoh:1985}
\bibinfo{author}{\bibfnamefont{H.}~\bibnamefont{{Munakata}}},
  \bibinfo{author}{\bibfnamefont{Y.}~\bibnamefont{{Kohyama}}},
  \bibnamefont{and} \bibinfo{author}{\bibfnamefont{N.}~\bibnamefont{{Itoh}}},
  \bibinfo{journal}{\apj} \textbf{\bibinfo{volume}{296}}, \bibinfo{pages}{197}
  (\bibinfo{year}{1985}).

\bibitem[{\citenamefont{{Kohyama} et~al.}(1986)\citenamefont{{Kohyama}, {Itoh},
  and {Munakata}}}]{Itoh:1986}
\bibinfo{author}{\bibfnamefont{Y.}~\bibnamefont{{Kohyama}}},
  \bibinfo{author}{\bibfnamefont{N.}~\bibnamefont{{Itoh}}}, \bibnamefont{and}
  \bibinfo{author}{\bibfnamefont{H.}~\bibnamefont{{Munakata}}},
  \bibinfo{journal}{\apj} \textbf{\bibinfo{volume}{310}}, \bibinfo{pages}{815}
  (\bibinfo{year}{1986}).

\bibitem[{\citenamefont{{Itoh} et~al.}(1989)\citenamefont{{Itoh}, {Adachi},
  {Nakagawa}, {Kohyama}, and {Munakata}}}]{Itoh:1989}
\bibinfo{author}{\bibfnamefont{N.}~\bibnamefont{{Itoh}}},
  \bibinfo{author}{\bibfnamefont{T.}~\bibnamefont{{Adachi}}},
  \bibinfo{author}{\bibfnamefont{M.}~\bibnamefont{{Nakagawa}}},
  \bibinfo{author}{\bibfnamefont{Y.}~\bibnamefont{{Kohyama}}},
  \bibnamefont{and}
  \bibinfo{author}{\bibfnamefont{H.}~\bibnamefont{{Munakata}}},
  \bibinfo{journal}{\apj} \textbf{\bibinfo{volume}{339}}, \bibinfo{pages}{354}
  (\bibinfo{year}{1989}).

\bibitem[{\citenamefont{{Itoh} et~al.}(1992)\citenamefont{{Itoh}, {Mutoh},
  {Hikita}, and {Kohyama}}}]{Itoh:1992}
\bibinfo{author}{\bibfnamefont{N.}~\bibnamefont{{Itoh}}},
  \bibinfo{author}{\bibfnamefont{H.}~\bibnamefont{{Mutoh}}},
  \bibinfo{author}{\bibfnamefont{A.}~\bibnamefont{{Hikita}}}, \bibnamefont{and}
  \bibinfo{author}{\bibfnamefont{Y.}~\bibnamefont{{Kohyama}}},
  \bibinfo{journal}{\apj} \textbf{\bibinfo{volume}{395}}, \bibinfo{pages}{622}
  (\bibinfo{year}{1992}).

\bibitem[{\citenamefont{{Itoh} et~al.}(1996)\citenamefont{{Itoh}, {Hayashi},
  {Nishikawa}, and {Kohyama}}}]{Itoh:1996}
\bibinfo{author}{\bibfnamefont{N.}~\bibnamefont{{Itoh}}},
  \bibinfo{author}{\bibfnamefont{H.}~\bibnamefont{{Hayashi}}},
  \bibinfo{author}{\bibfnamefont{A.}~\bibnamefont{{Nishikawa}}},
  \bibnamefont{and}
  \bibinfo{author}{\bibfnamefont{Y.}~\bibnamefont{{Kohyama}}},
  \bibinfo{journal}{\apj} \textbf{\bibinfo{volume}{102}}, \bibinfo{pages}{411}
  (\bibinfo{year}{1996}).

\bibitem[{\citenamefont{{Haft} et~al.}(1994)\citenamefont{{Haft}, {Raffelt},
  and {Weiss}}}]{Haft:1994}
\bibinfo{author}{\bibfnamefont{M.}~\bibnamefont{{Haft}}},
  \bibinfo{author}{\bibfnamefont{G.}~\bibnamefont{{Raffelt}}},
  \bibnamefont{and} \bibinfo{author}{\bibfnamefont{A.}~\bibnamefont{{Weiss}}},
  \bibinfo{journal}{\apj} \textbf{\bibinfo{volume}{425}}, \bibinfo{pages}{222}
  (\bibinfo{year}{1994}), \eprint{astro-ph/9309014}.

\bibitem[{\citenamefont{Woosley and Heger}(2007)}]{WOOSLEY:2007}
\bibinfo{author}{\bibfnamefont{S.}~\bibnamefont{Woosley}} \bibnamefont{and}
  \bibinfo{author}{\bibfnamefont{A.}~\bibnamefont{Heger}},
  \bibinfo{journal}{Physics Reports} \textbf{\bibinfo{volume}{442}},
  \bibinfo{pages}{269 } (\bibinfo{year}{2007}), ISSN \bibinfo{issn}{0370-1573},
  \bibinfo{note}{the Hans Bethe Centennial Volume 1906-2006}.

\bibitem[{\citenamefont{Izaguirre et~al.}(2013)\citenamefont{Izaguirre,
  Krnjaic, Schuster, and Toro}}]{Izaguirre:2013}
\bibinfo{author}{\bibfnamefont{E.}~\bibnamefont{Izaguirre}},
  \bibinfo{author}{\bibfnamefont{G.}~\bibnamefont{Krnjaic}},
  \bibinfo{author}{\bibfnamefont{P.}~\bibnamefont{Schuster}}, \bibnamefont{and}
  \bibinfo{author}{\bibfnamefont{N.}~\bibnamefont{Toro}},
  \bibinfo{journal}{Phys. Rev.} \textbf{\bibinfo{volume}{D88}},
  \bibinfo{pages}{114015} (\bibinfo{year}{2013}), \eprint{1307.6554}.

\bibitem[{\citenamefont{Batell et~al.}(2014{\natexlab{b}})\citenamefont{Batell,
  Essig, and Surujon}}]{Batell:2014}
\bibinfo{author}{\bibfnamefont{B.}~\bibnamefont{Batell}},
  \bibinfo{author}{\bibfnamefont{R.}~\bibnamefont{Essig}}, \bibnamefont{and}
  \bibinfo{author}{\bibfnamefont{Z.}~\bibnamefont{Surujon}},
  \bibinfo{journal}{Phys. Rev. Lett.} \textbf{\bibinfo{volume}{113}},
  \bibinfo{pages}{171802} (\bibinfo{year}{2014}{\natexlab{b}}),
  \eprint{1406.2698}.

\bibitem[{\citenamefont{Sung et~al.}(2019)\citenamefont{Sung, Tu, and
  Wu}}]{Sung2019}
\bibinfo{author}{\bibfnamefont{A.}~\bibnamefont{Sung}},
  \bibinfo{author}{\bibfnamefont{H.}~\bibnamefont{Tu}}, \bibnamefont{and}
  \bibinfo{author}{\bibfnamefont{M.-R.} \bibnamefont{Wu}}
  (\bibinfo{year}{2019}), \eprint{1903.07923}.

\bibitem[{\citenamefont{Weldon}(1982)}]{Weldon:1982}
\bibinfo{author}{\bibfnamefont{H.~A.} \bibnamefont{Weldon}},
  \bibinfo{journal}{Phys. Rev. D} \textbf{\bibinfo{volume}{26}},
  \bibinfo{pages}{1394} (\bibinfo{year}{1982}).

\end{thebibliography}
\end{document}